\documentclass{article}
\usepackage[utf8]{inputenc}
\usepackage{geometry}
\usepackage{amsmath}
\usepackage{amsfonts}
\usepackage{booktabs}
\usepackage[pdftex]{graphicx} 
\usepackage{color}
\title{SUNDIALS Multiphysics+\texttt{MPIManyVector} Performance Testing}
\author{Daniel R. Reynolds, David J. Gardner, Cody J. Balos, \\
and Carol S. Woodward}
\date{September 2019}

\newcommand{\bx}{\mathbf{x}}
\newcommand{\bc}{\mathbf{c}}
\newcommand{\bw}{\mathbf{w}}
\newcommand{\by}{\mathbf{y}}
\newcommand{\bz}{\mathbf{z}}
\newcommand{\bv}{\mathbf{v}}
\newcommand{\bzeta}{\mathbf{\zeta}}
\newcommand{\br}{\mathbf{r}}
\newcommand{\bF}{\mathbf{F}}
\newcommand{\bR}{\mathbf{R}}
\newcommand{\bG}{\mathbf{G}}
\newcommand{\R}{\mathbb{R}}
\newcommand{\bfs}{\mathbf{f}^{S}}
\newcommand{\bff}{\mathbf{f}^{F}}
\newcommand{\bfrhs}{\mathbf{f}}
\newcommand{\bFres}{\mathbf{\mathcal{F}}}
\newcommand{\Hy}{\operatorname{H}}
\newcommand{\He}{\operatorname{He}}
\newcommand{\el}{\operatorname{e}}

\begin{document}

\thispagestyle{empty}
\begin{figure}[tbph]
  \vspace*{-1.5cm}
  \centerline{
    \includegraphics[width=1.2\linewidth]{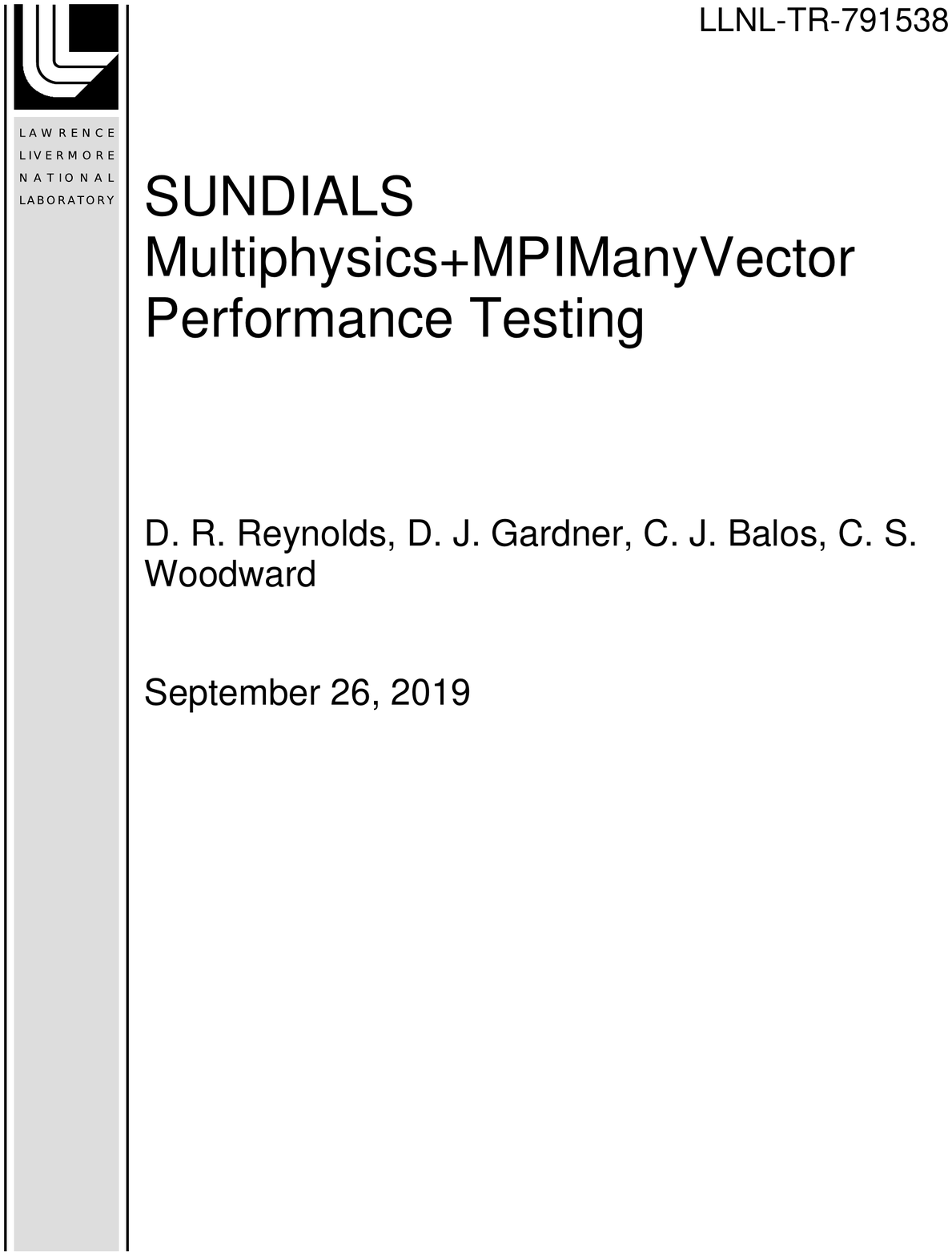}
  }
\end{figure}

\clearpage

\thispagestyle{empty}
\begin{figure}[tbph]
  \vspace*{-1.5cm}
  \centerline{
    \includegraphics[width=1.2\linewidth]{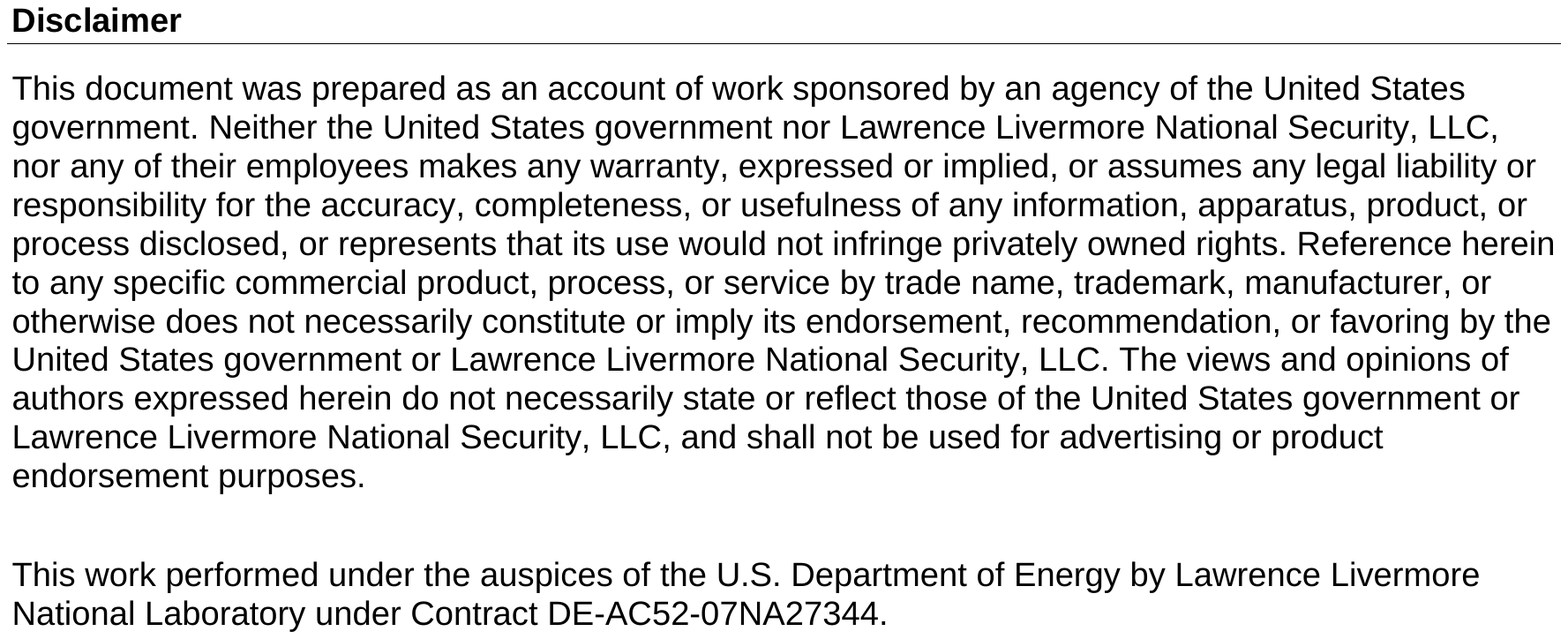}
  }
\end{figure}

\clearpage

\setcounter{page}{1}

\maketitle

\section{Introduction}
\label{sec:intro}

In this report we document performance test results on a
SUNDIALS-based multiphysics demonstration application.  We aim to 
assess the large-scale parallel performance of new
capabilities that have been added to the SUNDIALS \cite{hindmarsh2005sundials,SUNDIALS_site} suite of time integrators and nonlinear solvers in recent years under
funding from both the Exascale Computing Project (ECP) and the
Scientific Discovery through Advanced Scientific (SciDAC) program, 
specifically:
\begin{enumerate}
\item SUNDIALS' new \texttt{MPIManyVector} module, that allows extreme
  flexibility in how a solution ``vector'' is staged on computational
  resources.
\item ARKode's new multirate integration module, \texttt{MRIStep},
  allowing high-order accurate calculations that subcycle ``fast''
  processes within ``slow'' ones. 
\item SUNDIALS' new flexible linear solver interfaces, that allow
  streamlined specification of problem-specific linear solvers. 
\item SUNDIALS' new \texttt{N\_Vector} additions of ``fused'' vector
  operations (to increase arithmetic intensity) and separation of 
  reduction operations into ``local'' and ``global'' versions (to reduce
  latency by combining multiple reductions into a single 
  \texttt{MPI\_Allreduce} call). 
\end{enumerate}
We anticipate that subsequent reports will extend this work to
investigate a variety of other new features, including SUNDIALS' 
generic \texttt{SUNNonlinearSolver} interface and accelerator-enabled 
\texttt{N\_Vector} modules, and upcoming \texttt{MRIStep}
extensions to support custom ``fast'' integrators (that leverage
problem structure) and IMEX integration of the ``slow'' time scale (to
add diffusion).

\section{Problem Description}
\label{sec:problem}

We simulate the three-dimensional nonlinear inviscid compressible Euler
equations, combined with advection and reaction of chemical species,
\begin{equation}
\label{eq:adv_rx_pde}
  \bw_t = -\nabla\cdot \bF(\bw) + \bR(\bw) + \bG(\bx,t).
\end{equation}
Here, the independent variables are $(\bx,t) = (x,y,z,t) \in \Omega \times (t_0,
t_f]$, where the spatial domain is a three-dimensional cube, $\Omega =
[x_l, x_r] \times [y_l, y_r] \times [z_l,z_r]$.  The partial differential
equation is completed using initial condition $\bw(\bx,t_0) = \bw_0(\bx)$ and
face-specific boundary conditions, \{xlbc,\, xrbc,\, ylbc,\, yrbc,\, zlbc,\,
zrbc\}, corresponding to conditions that may be separately applied at
the spatial locations $\{(x_l,y,z)$, $(x_r,y,z)$, $(x,y_l,z)$,
$(x,y_r,z)$, $(x,y,z_l)$, $(x,y,z_r)\}$, respectively, and where each
condition may be any one of:
\begin{itemize}
\item periodic (requires that \emph{both} faces in this direction use
  this condition),
\item homogeneous Neumann (i.e., $\nabla w_i(\bx,t)\cdot \mathrm{n}=0$
  for $\bx\in\partial\Omega$ with outward-normal vector $\mathbf{n}$, 
  and for each species $w_i$), 
\item homogeneous Dirichlet (i.e., $w_i(\bx,t) = 0$ for
  $\bx\in\partial\Omega$ and for each species $w_i$), or 
\item reflecting (i.e., homogeneous Neumann for all species
  \emph{except} the momentum field perpendicular to that face, that
  has a homogeneous Dirichlet condition), 
\end{itemize}
The computed solution is given by
$\bw = \begin{bmatrix} \rho & \rho v_x & \rho v_y & \rho v_z & e_t & \bc \end{bmatrix}^T
= \begin{bmatrix} \rho & m_x & m_y & m_z & e_t & \bc\end{bmatrix}^T$,
that corresponds to the density $(\rho)$, $x$,$y$,$z$-momentum
$(m_x,m_y,m_z)$, total energy per unit volume $(e_t)$, and vector of
chemical densities $(\bc \in\R^{n_c})$ that are advected
along with the fluid.  The advective fluxes $\bF = \begin{bmatrix}F_x
  & F_y & F_z\end{bmatrix}^T$ are given by 
\begin{align}
  \label{eq:xflux}
  F_x(\bw) &= \begin{bmatrix} \rho v_x & \rho v_x^2 + p & \rho v_x v_y &
    \rho v_x v_z & v_x (e_t+p) & \bc v_x \end{bmatrix}^T\\ 
  \label{eq:yflux}
  F_y(\bw) &= \begin{bmatrix} \rho v_y & \rho v_x v_y & \rho v_y^2 + p &
    \rho v_y v_z & v_y (e_t+p) & \bc v_y \end{bmatrix}^T\\ 
  \label{eq:zflux}
  F_z(\bw) &= \begin{bmatrix} \rho v_z & \rho v_x v_z & \rho v_y v_z &
    \rho v_z^2 + p & v_z (e_t+p) & \bc v_z \end{bmatrix}^T. 
\end{align}
The reaction term $\bR(\bw)$ and external force $\bG(\bx,t)$ are
test-problem-dependent, and the ideal gas equation of state relates the pressure
and total energy density,
\begin{align}
  \notag
  p &= \frac{R}{c_v}\left(e_t - \frac{\rho}{2} (v_x^2 + v_y^2 + v_z^2)\right)\\
  \label{eq:eos}
  \Leftrightarrow\qquad&\\
  \notag
  e_t &= \frac{p c_v}{R} + \frac{\rho}{2}(v_x^2 + v_y^2 + v_z^2),
\end{align}
or equivalently,
\begin{align}
  \notag
  p &= (\gamma-1) \left(e_t - \frac{\rho}{2} (v_x^2 + v_y^2 + v_z^2)\right)\\
  \label{eq:eos2}
  \Leftrightarrow\qquad&\\
  \notag
  e_t &= \frac{p}{\gamma-1} + \frac{\rho}{2}(v_x^2 + v_y^2 + v_z^2),
\end{align}
The above model includes the physical parameters:
\begin{itemize}
\item $R$ is the specific ideal gas constant (287.14 J/kg/K for air).
\item $c_v$ is the specific heat capacity at constant volume (717.5
  J/kg/K for air),
\item $\gamma$ is the ratio of specific heats, $\gamma = \frac{c_p}{c_v} =
  1 + \frac{R}{c_v}$; this is typically 1.4 for air, and $5/3$ for astrophysical
  gases.
\end{itemize}
The speed of sound in the gas is given by
\begin{equation}
  \label{eq:soundspeed}
  c = \sqrt{\frac{\gamma p}{\rho}}.
\end{equation}
The fluid variables ($\rho$, $\mathbf{m}$, and $e_t$) are non-dimensionalized; 
when converted to physical CGS values these have units:
\begin{itemize}
\item $[\rho] = \text{g}/\text{cm}^3$,
\item $[v_x] = [v_y] = [v_z] = \text{cm}/\text{s}$, which implies that 
  $[m_x] = [m_y] = [m_z] = \text{g}/\text{cm}^2/\text{s}$,
\item  $[e_t] = \text{g}/\text{cm}/\text{s}^2$.
\end{itemize}
The chemical densities have physical units $[\bc_i] = \text{g}/\text{cm}^3$, although when
these are transported by the fluid these are converted to dimensionless units as well.

\section{Implementation}
\label{sec:implementation}

We apply a method of lines approach for converting the system of partial 
differential equations (PDEs) \eqref{eq:adv_rx_pde} into a discrete set 
of equations.  To this end, we first discretize in space, converting the 
PDE system into a very large system of ordinary differential equation 
(ODE) initial-value problems (IVPs).  We then apply the \texttt{MRIStep} 
time-integrator from the ARKode SUNDIALS package.  This time 
integration approach in turn requires a sub-integrator for the ``fast'' 
chemical reactions, for which we employ the \texttt{ARKStep} 
time-integrator, also from ARKode.  \texttt{ARKStep}, in turn, 
requires the solution of very large-scale systems of nonlinear 
algebraic equations.  We discuss our use of each of the above components
in the following subsections.

\subsection{Spatial discretization}
\label{sec:implementation:space}

\begin{figure}[tbph]
  \centerline{
    \includegraphics[width=.9\linewidth]{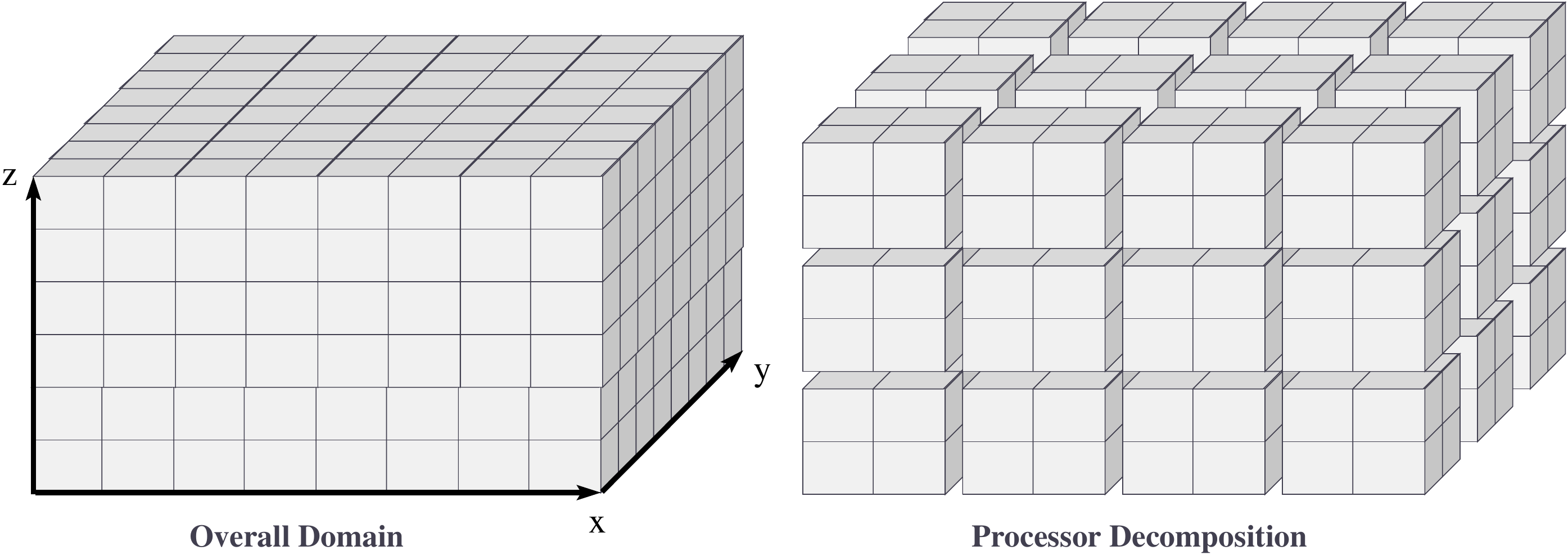}
  }
  \caption{Illustration of 3-dimensional domain decomposition
    algorithm with 36 MPI tasks broken into a $4\times
    3\times 3$ layout.  Each MPI task in this illustration owns
    a $2\times 2\times 2$ local grid.} 
  \label{fig:3d-decomp}
\end{figure}

We discretize the domain $\Omega$ into a uniform grid of
dimensions $n_x \times n_y \times n_z$, such that we have a three-dimensional
rectangular cuboid of cell-centered values $(x_i,y_j,z_k)$ wherein
\begin{align*}
  x_i = x_l + \left(i + \frac12\right) \Delta x, &\qquad \Delta x = \frac{x_r-x_l}{n_x}, \qquad i=0,\ldots,n_x-1\\
  y_j = y_l + \left(j + \frac12\right) \Delta y, &\qquad \Delta y = \frac{y_r-y_l}{n_y}, \qquad j=0,\ldots,n_y-1,\\
  z_k = z_l + \left(k + \frac12\right) \Delta z, &\qquad \Delta z = \frac{z_r-z_l}{n_z}, \qquad k=0,\ldots,n_z-1.
\end{align*}
This spatial domain is then decomposed in parallel using a standard 3D
domain decomposition approach over $n_p$ MPI tasks, with layout
$n_{px} \times n_{py} \times n_{pz}$, defined automatically via the
\texttt{MPI\_Dims\_create} utility routine, as illustrated in Figure
\ref{fig:3d-decomp}.  This results in each MPI task ``owning'' a
local grid of dimensions $n_{xloc} \times n_{yloc} \times n_{zloc}$.

\begin{figure}[tbph]
  \centerline{
    \includegraphics[width=.9\linewidth]{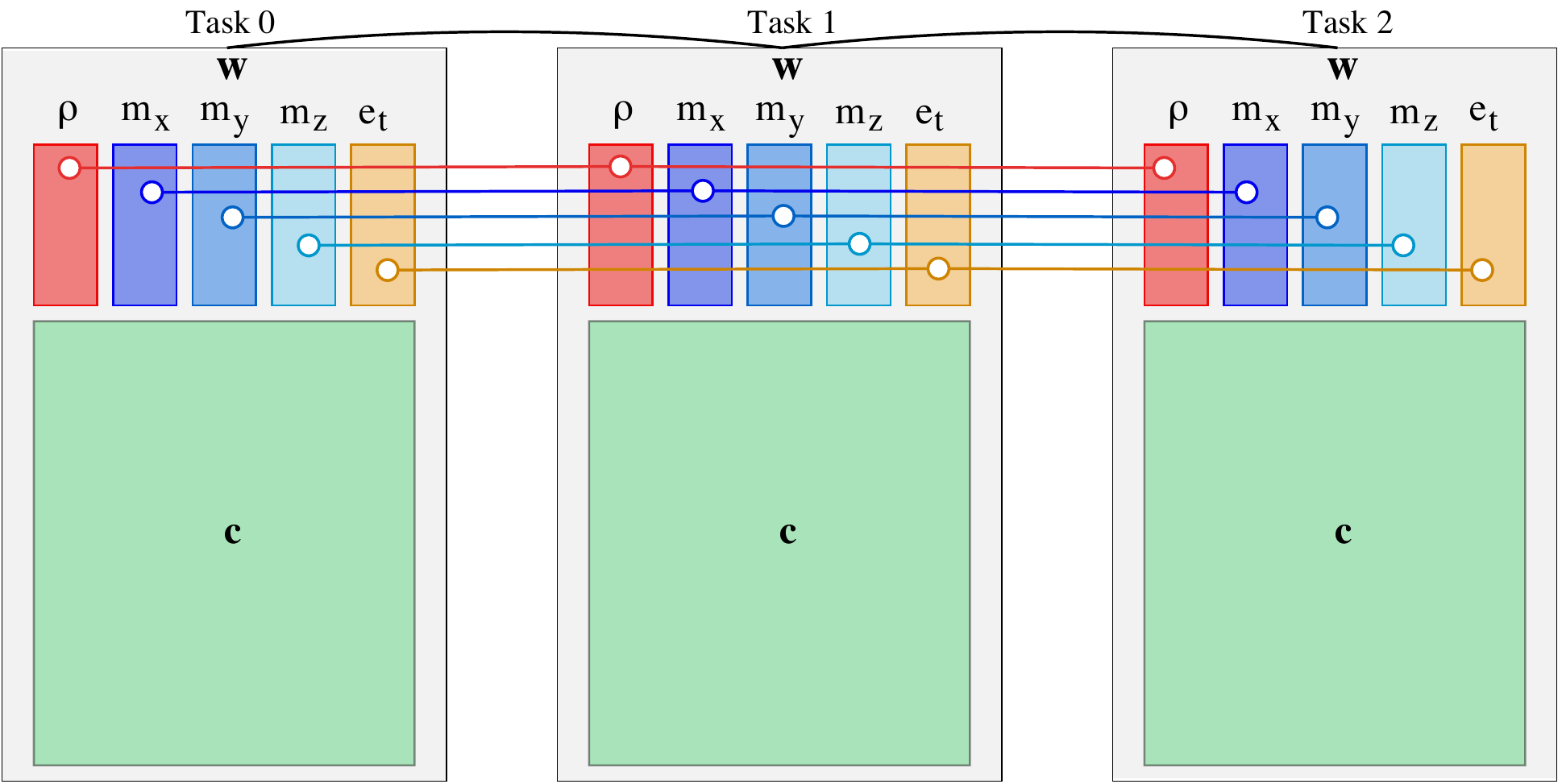}
  }
  \caption{Illustration of \texttt{MPIManyVector} use in this problem.
    Each of the fluid fields ($\rho$, $m_x$, $m_y$, $m_z$ and $e_t$) are
    stored in \texttt{N\_Vector\_Parallel} objects, connected via the
    3-dimensional Cartesian MPI communicator.  The chemical densities,
    however, are stored together in a single
    \texttt{N\_Vector\_Serial} object on each MPI task (with no
    communicator directly connecting them).  These six vectors on each
    MPI task are then grouped together into a single
    \texttt{N\_Vector\_MPIManyVector}, $\bw$, that inherits the
    3-dimensional Cartesian MPI communicator from its parallel
    subvectors.}
  \label{fig:manyvector}
\end{figure}

Within each cell in the domain we store $5+n_c$ variables,
corresponding to the values of $\bw$ at that spatial location.  Here,
we employ the newly-introduced \texttt{N\_Vector\_MPIManyVector}
implementation, that allows creation of a single \texttt{N\_Vector}
out of any valid set of subsidiary \texttt{N\_Vector} objects.  To this
end, we store each of the five fluid fields ($\rho$, $m_x$, $m_y$,
$m_z$ and $e_t$) in its own \texttt{N\_Vector\_Parallel} object to
simplify access and I/O.  We then store all chemical species owned by each
MPI task in a single \texttt{N\_Vector\_Serial} object
(\emph{note: this will eventually be changed to use a device-specific
  \texttt{N\_Vector} object, such as \texttt{N\_Vector\_CUDA},
  \texttt{N\_Vector\_RAJA}, or \texttt{N\_Vector\_OpenMPDEV}}).  Each
MPI task then combines its pointers for the five fluid vectors, along
with its own chemical species vector, into its full ``solution''
\texttt{N\_Vector\_MPIManyVector}, $\bw$, using the
\texttt{N\_VNew\_MPIManyVector} routine.  An illustration of this
\texttt{MPIManyVector} structure is shown in Figure \ref{fig:manyvector}.


The final item to note with regard to the spatial discretization is
our approach for approximating the flux divergence $\nabla\cdot \bF(\bw)$
shown in equation \eqref{eq:adv_rx_pde}.  For this, we apply a
5th-order FD-WENO reconstruction, where we precisely follow the
algorithm laid out in the seminal paper by Shu \cite{Shu2003}, which
we briefly summarize here.  In order to properly conserve mass,
momentum, and energy in \eqref{eq:adv_rx_pde}, we first compute the
fluxes at each of the six faces surrounding the cell $(x_i,y_j,z_k)$,
and apply these in a standard conservative fashion, namely
\begin{align}
  \notag
  \nabla\cdot \bF(\bw(x_i,y_j,z_k,t))
  \approx 
  &\frac{1}{\Delta x}\left[ F_x(\bw(x_{i+1/2},y_j,z_k,t)) - F_x(\bw(x_{i-1/2},y_j,z_k,t))\right] + \\
  \label{eq:div_flux}
  &\frac{1}{\Delta y}\left[ F_y(\bw(x_i,y_{j+1/2},z_k,t)) - F_x(\bw(x_i,y_{j-1/2},z_k,t))\right] + \\
  \notag
  &\frac{1}{\Delta z}\left[ F_z(\bw(x_i,y_j,z_{k+1/2},t)) - F_x(\bw(x_i,y_j,z_{k-1/2},t))\right].
\end{align}
Since the cells $(x_i,y_j,z_k)$ and $(x_{i+1},y_j,z_k)$ share the same
flux value $F_x(\bw(x_{i+1/2},y_j,z_k,t))$, this results in
conservation to full machine precision (modulo boundary conditions,
source terms, and reaction processes).  The 5th-order FD-WENO scheme is
used to construct each of these face-centered flux values.  This algorithm 
computes the flux $F_x(\bw(x_{i+1/2},y_j,z_k,t))$ using a 6-point stencil of 
solution values $\bw(x_{i-2},y_j,z_k,t)$, $\bw(x_{i-1},y_j,z_k,t)$,
$\bw(x_{i},y_j,z_k,t)$, $\bw(x_{i+1},y_j,z_k,t)$,
$\bw(x_{i+2},y_j,z_k,t)$, and $\bw(x_{i+3},y_j,z_k,t)$, i.e., the
6 ``closest'' cell centers to the face $(x_{i+1/2},y_j,z_k)$ along the
$x$-direction.  The stencils are analagous in the $y$ and $z$ directions.  
Thus under our three-dimensional domain decomposition approach
outlined above, each MPI task must obtain three layers of ``ghost'' cells from
each of its six neighboring subdomains.

We thus compute the flux divergence $\nabla\cdot \bF(\bw)$ using the
following steps:
\begin{enumerate}
\item Begin exchange of boundary layers with neighbors via
  asynchronous \texttt{MPI\_Isend} and \texttt{MPI\_Irecv} calls.
\item Compute and store the fluxes at each face in the strict interior
  of the subdomain, i.e., for all cell faces whose 6-point stencil
  involves no data from neighboring MPI tasks.  For example, we compute
  $F_x(\bw(x_{i-1/2},y_j,z_k))$ over the ranges $i=3,\ldots,n_{xloc}-3$.
  Since the vast majority of computational effort in this portion of the
  algorithm lies within the arithmetically intense FD-WENO 
  reconstruction itself, we first copy the local stencil of 
  $6 \times (5 + n_c)$ unknowns from their various \texttt{N\_Vector} 
  locations into a contiguous buffer before performing the FD-WENO
  reconstruction at each face.
\item Wait for completion of all asynchronous \texttt{MPI\_Isend} and
  \texttt{MPI\_Irecv} calls.
\item Compute the face-centered fluxes near subdomain boundaries, i.e., 
  those face-valued fluxes that were omitted above.  As these stencils now 
  depend on values from neighboring subdomains, the only difference 
  from step 2 is that this must use a different routine to copy
  the local stencil into the contiguous buffer, since these copies must 
  appropriately handle data from the \texttt{MPI\_Irecv} buffers.
\item Finally, compute the formula \eqref{eq:div_flux} over the entire
  local subdomain, i.e., for each location $(x_i,y_j,z_k)$ over the
  ranges $i=0,\ldots,n_{xloc}-1$, $j=0,\ldots,n_{yloc}-1$, and
  $k=0,\ldots,n_{zloc}-1$.
\end{enumerate}
We note that due to the high arithmetic intensity of the FD-WENO approach, 
each MPI task need not own a very large computational subdomain 
for the point-to-point communication to be completely overlaid by 
computation in step 2.

\subsection{Temporal discretization}
\label{sec:implementation:time}

After spatial semi-discretization, the original PDE system
\eqref{eq:adv_rx_pde} may be written as an ODE initial-value problem,
\begin{equation}
  \label{eq:ivp}
  \by_t = \bfs(t,\by) + \bff(t,\by),
\end{equation}
where $\by$ contains the spatial semi-discretization of the solution vector 
$\bw$, $\bfs(t,\by)$ contains the spatial semi-discretization of the
terms $\left(\bG(\bx,t)-\nabla\cdot \bF(\bw)\right)$ and $\bff(t,\by)$
contains the spatial semi-discretization of the term $\bR(\bw)$.
Here, we use the ``$S$'' superscript to denote the ``slow'' dynamical processes
(advection and externally-applied forces), and ``$F$'' to denote the
``fast'' reaction processes.

For non-reactive flows in which $\bff=\mathbf{0}$, the initial value
problem is nonstiff, and is therefore solved using a
temporally-adaptive explicit Runge--Kutta method from ARKode's
\texttt{ARKStep} module.  While that use case is supported by our
demonstration code, we do not examine this ``single physics'' use case
here.

For problems involving chemical reactions, the multiphysics
initial-value problem \eqref{eq:ivp} typically exhibits multiple
temporal scales.  We therefore solve these problems using ARKode's
\texttt{MRIStep} module, that employs a third-order accurate
\emph{multirate infinitesimal step} method for problems characterized by
two time scales \cite{Schlegel2009,Schlegel2012a,Schlegel2012b}.  
Here, a single time
step to evolve $\by(t_{n-1}) \to \by(t_{n-1}+h^S)$, denoted by
$\by_{n-1}\to\by_{n}$, for the full initial-value problem
\eqref{eq:ivp}, proceeds according to the following algorithm:

\begin{enumerate}
\item set $\bz_1 = \by_{n-1}$,
\item for $i = 2,\ldots,s+1$:
\begin{enumerate}
\item define the ``fast'' initial condition: $\bv(t^S_{n,i-1}) = \bz_{i-1}$,
\item compute the forcing term
  \[
    \br = \frac{1}{c^S_i - c^S_{i-1}} \sum_{j=1}^{i-1}
    (A^S_{i,j} - A^S_{i-1,j})\, \bfs(t^S_{n,j}, \bz_j),
  \]
\item for $\tau \in (t^S_{n,i-1}, t^S_{n,i}]$, solve the ``fast''
  initial-value problem
  \begin{equation}
    \label{eq:ivp_fast}
    \dot{\bv}(\tau) = \bff(\tau, \bv) + \br,
  \end{equation}
\item set the new ``slow'' stage: $\bz_i = \bv(t^S_{n,i})$,
\end{enumerate}
\item set the time-evolved solution: $\by_{n} = \bz_{s+1}$,
\end{enumerate}
where $(A^S, b^S, c^S)$ correspond to the coefficients for an
explicit, $s$-stage ``slow'' Runge--Kutta method, $A^S$ is
padded with a final row, $A^S_{s+1,j}=b^S_{j}$, and where 
$t^S_{n,j} = t_{n-1} + c^S_{j} h^S$ for $j=1,\ldots,s$ correspond to 
the ``slow'' stage times.  In this demonstration application, we use 
the default ``KW3'' \texttt{MRIStep} slow Runge--Kutta coefficients 
\cite{Knoth1998}.  For evolution of the ``fast'' problems 
\eqref{eq:ivp_fast} above, we use a temporally-adaptive, 
diagonally-implicit Runge--Kutta method from ARKode's 
\texttt{ARKStep} module, namely the ``ARK437L2SA'' DIRK
method from \cite{Kennedy2019}.

As we will describe in Section \ref{sec:test_problem} below when 
discussing our physical test problem, chemically-reactive flows exhibit 
fast transient behavior, so their stable evolution heavily depends on the
inherent robustness of temporally-adaptive integration.  However, the
primary purpose of this report is to document parallel performance of
the \texttt{MPIManyVector} and other algebraic solver enhancements
that have been recently added to SUNDIALS.  As such, we employ a
hybrid adaptive + fixed-step integration approach for the fast
time-scale subproblems \eqref{eq:ivp_fast}.  Specifically, we
partition the overall temporal domain $(t_0,t_f]$ into two parts,
$(t_0,t_0+h_t]$ and $(t_0+h_t,t_f]$.  The first portion is considered
the ``transient'' time period, where chemical species exhibit
\emph{very} fast dynamical changes, as they rapidly adjust from their
initial conditions to their slower (but still fast) solution
trajectories.  This time period is therefore evolved with
\texttt{ARKStep}'s temporal adaptivity enabled; the adaptivity
parameters employed for this phase of the simulations are provided in
Table \ref{tab:TransientCoefficients}.
\begin{table}[!htp]
  \begin{center}
    \begin{tabular}{|l|l|}
      \hline
      Parameter & Value \\
      \hline
      \texttt{ARKStepSetAdaptivityMethod} & $(2, 1, 0)$\\
      \texttt{ARKStepSetMaxNumSteps} & 5000\\
      \texttt{ARKStepSetSafetyFactor} & 0.99\\
      \texttt{ARKStepSetErrorBias} & 2.0\\
      \texttt{ARKStepSetMaxGrowth} & 2.0\\
      \texttt{ARKStepSetMaxNonlinIters} & 10\\
      \texttt{ARKStepSetNonlinConvCoef} & 0.01\\
      \texttt{ARKStepSetMaxStep}$^1$ & $h^S/1000$\\
      Relative tolerance & $10^{-5}$\\
      Absolute tolerance & $10^{-9}$\\
      \hline
    \end{tabular}
    \caption{\texttt{ARKStep} parameters for adaptive integration of
      initial transient dynamics.  Only values that are changed from
      the standard \texttt{ARKStep} defaults are shown.  For the
      line \texttt{ARKStepSetMaxStep}, $h^S$ is the value of the fixed
      step size that was used to evolve the ``slow'' dynamics -- due
      to the explicit CFL stability condition, this is adjusted in
      proportion to the spatial mesh size, i.e., $h^S \propto
      \min(\Delta x, \Delta y, \Delta z)$.
      \label{tab:TransientCoefficients}
    }
  \end{center}
\end{table}

The second (and typically much longer) time interval, $(t_0+h_t,t_f]$,
is evolved using a fixed ``fast'' time step size, $h^F = h^S/1000$,
thereby bypassing \texttt{ARKStep}'s built-in temporal adaptivity
approaches to produce a more predictable amount of work as the problem
is pushed to larger scales.  The challenge with using fixed time step 
sizes in such an application is that in fixed-step mode, any algebraic 
solver convergence failure becomes fatal, and causes the
entire simulation to halt.  Thus, the value of $h_t$ must be chosen
appropriately, to balance the need for adaptivity-based robustness
during initial transient chemical evolution (i.e., larger $h_t$)
against the desire for a fixed amount of computational work
per MPI task when performing weak scaling studies (i.e., smaller
$h_t$).  We present the values used in the current studies in Section 
\ref{sec:test_problem}, when discussing the particular test problem 
used here.

Both evolution periods, $(t_0,t_0+h_t]$ and $(t_0+h_t,t_f]$, are
evolved using single calls to \texttt{MRIStepEvolve}, although
the second period can be broken into a sequence of separate subperiods
so that solution statistics can be displayed,  and/or solution
checkpoint files can be written to disk.  However, since this study
focuses on overall solver performance, all such diagnostic and
solution output is disabled.

\subsection{Algebraic solvers}
\label{sec:implementation:algebraic}

Since the slow time scale is currently treated explicitly, the only
algebraic solvers present in these calculations occur when evolving
each ``fast'' subproblem \eqref{eq:ivp_fast}.  As these subproblems
are evolved using diagonally implicit Runge--Kutta (DIRK) methods,
each stage may require the solution of a nonlinear algebraic system of 
equations.  We briefly outline the structure of these DIRK methods,
and then discuss the algebraic solvers used for these problems.

Considering the fast IVP
\[
  \dot{\bv}(\tau) = \bff(\tau, \bv) + \br \equiv \bfrhs(\tau,\bv),
\]
a $\sigma$-stage DIRK method with coefficients $(A,b,c)$ evolves one
time step $\bv(\tau_{m-1}) \to \bv(\tau_{m-1}+\theta)$, denoted for
short by $\bv_{m-1}\to\bv_{m}$, via the algorithm: 
\begin{enumerate}
\item for $i = 1,\ldots,\sigma$, solve for the ``stages'' $\bzeta_i$ that satisfy the equations
  \begin{equation}
    \label{eq:DIRK_stages}
    \bzeta_i = \bv_{m-1} + \theta \sum_{j=1}^{i} A_{i,j} \bfrhs(\tau_{m,j}, \bzeta_j),
  \end{equation}
  where $\tau_{m,j} = \tau_{m-1} + c_j \theta$,
\item compute the time-evolved solution
  \[
    \bv_m = \bv_{m-1} + \theta \sum_{i=1}^{\sigma} b_i \bfrhs(\tau_{m,i}, \bzeta_i),
  \]
\item (optional) compute the embedded solution
  \[
    \tilde{\bv}_m = \bv_{m-1} + \theta \sum_{i=1}^{\sigma} 
    \tilde{b}_i \bfrhs(\tau_{m,i}, \bzeta_i).
  \]
\end{enumerate}

The solution of at most $\sigma$ nonlinear algebraic systems \eqref{eq:DIRK_stages}
is required to compute each stage $\bzeta_i$. Writing these systems in standard
root-finding form, for each fast substep we must solve $\sigma$
separate nonlinear algebraic systems:
\begin{align}
  \label{eq:residual}
  0 = \bFres(\bzeta_i)
  \equiv \Bigg[\bzeta_i - \theta A_{i,i}
    \bfrhs(\tau_{m,i},\bzeta_i)\Bigg] - \left[\bv_{m-1} + \theta
    \sum_{j=1}^{i-1} A_{i,j} \bfrhs(\tau_{m,j},\bzeta_j)\right],
    \qquad i=1,\ldots,\sigma.
\end{align}
where the first bracketed term contains the implicit portions of the
nonlinear residual, and the second bracketed term contains known
data.  We note that each $\bzeta_i \in \R^{n_x n_y n_z (5+n_c)}$ can be
a \emph{very} large vector.  However since $\bfrhs = \bff(\tau, \bv) +
\br$, and $\bff(\tau,\bv)$ is just the spatially semi-discretized
version of the reaction function $\bR(\bv)$, we see that although
equation \eqref{eq:residual} is nonlinear, it only involves couplings
between unknowns that are co-located at each spatial location, 
$(x_i,y_j,z_k)$.  Thus we may instead consider equation 
\eqref{eq:residual} to be equivalent to a system of $n_x n_y n_z$
\underline{separate} nonlinear systems of equations, each coupling only 
$5+n_c$ unknowns.

We may therefore leverage this structure in a myriad of ways to
improve parallel performance.  At one extreme, we may break
apart the large nonlinear system \eqref{eq:residual} into $n_x n_y
n_z$ separate nonlinear systems, performing an independent Newton
iteration separately at each spatial location.  At a coarser 
level, we could instead break apart equation \eqref{eq:residual} into
$n_{px} n_{py} n_{pz}$ separate nonlinear systems, one per MPI task, 
so that each Newton iteration may proceed without parallel 
communication.  The relative merits of these choices (as well as 
intermediate options, e.g., breaking this into two-dimensional 
``slabs'' or one-dimensional ``pencils'' of spatial cells) is not 
investigated here, but may be pursued in future efforts.  In this 
work, we utilize an even  coarser approach that solves the full 
nonlinear system of equations \eqref{eq:residual}
using a \emph{single} modified Newton iteration (the \texttt{ARKStep}
default), that couples all unknowns across the entire parallel
machine.  However, we exploit the problem structure by providing a
custom \emph{linear} solver module that solves each MPI task-local
linear system independently.  Due to the structure of $\bfrhs$, the
Jacobian matrix $J = \frac{\partial \bfrhs(\bzeta)}{\partial \bzeta}$
is block-diagonal,
\[
  J = \begin{bmatrix} J_1 & & & \\
  & J_2 & & \\
  & & \ddots & \\
  & & & J_{n_p} \end{bmatrix},
\]
where each MPI task-local block $J_p
\in\R^{n_{xloc}n_{yloc}n_{zloc}(5+n_c)}$, is itself block-diagonal,
\[
  J_p = \begin{bmatrix} J_{p,1,1,1} & & & \\
  & J_{p,2,1,1} & & \\
  & & \ddots & \\
  & & & J_{p,n_{xloc},n_{yloc},n_{zloc}} \end{bmatrix},
\]
with each cell-local block $J_{p,i,j,k} \in \R^{5+n_c}$.  We leverage
the extreme sparsity of these MPI task-local Jacobian matrices $J_p$ by
constructing each $J_p$ matrix in compressed-sparse-row (CSR) format, and storing
them using standard \texttt{SUNSparseMatrix} objects.  These are thus
automatically converted to Newton system matrices $A = I-\gamma J_p$
within \texttt{ARKStep}'s generic direct linear solver interface.  We
then leverage the block-diagonal structure and solve the overall
Newton linear systems $Ax=b$ using the \texttt{SUNLinSol\_KLU} linear
solver module separately on each MPI task.

As this is an optimally-parallel direct linear solver for the
block-diagonal Newton linear systems, and since $\bfrhs$ itself
involves no parallel communication, the only ``wasted'' MPI
communication that occurs in our solution of each nonlinear algebraic
system \eqref{eq:residual} is the computation of residual norms
$\|\bFres(\bzeta)\|_{WRMS}$ to determine completion of the Newton
iteration.  In future studies we plan to supply a custom nonlinear
solver that will remove this extraneous communication.  We also plan
to explore alternate strategies that will further leverage the
block-diagonal structure (slabs, pencils, etc.), particularly once we
transition simulation of the chemical kinetics to GPU accelerators.

\section{Test Description}
\label{sec:test_problem}

In this work, we consider the test problem of an advecting and
reacting primordial gas.  In addition to the five ``standard'' fluid
variables described in Section \ref{sec:problem}, we evolve 10
chemical species (i.e., $n_c=10$), for a total of 15 variables per
spatial location.  These species model the chemical behavior of a
low density primordial gas, present in models of the early universe
\cite{Abel1997,Glover2008,Kreckel2010,Trevisan_2002}.  This model consists of the species:
\begin{itemize}
\item $\Hy$ -- neutral atomic Hydrogen density
\item $\Hy^+$ -- positively-ionized atomic Hydrogen density
\item $\Hy^-$ -- negatively-ionized atomic Hydrogen density
\item $\Hy_2$ -- neutral molecular Hydrogen density
\item $\Hy_2^{+}$ -- ionized molecular Hydrogen density
\item $\He$ -- neutral atomic Helium density
\item $\He^{+}$ -- partially-ionized atomic Helium density
\item $\He^{++} $ -- fully-ionized atomic Helium density
\item $\el^-$ -- free electron density
\item $e_g$ -- internal gas energy (proportional to temperature)
\end{itemize}
The full set of chemical rate equations that encode these reactions
comprise the reaction term, $\bR(\bw)$, used in our code.  Both the
routine to evaluate this ``right-hand side'' function, as well as a
corresponding routine to evaluate its Jacobian in CSR format, are
provided by the Dengo software package \cite{Dengo_site}, a
source-code generation utility that translates from astropysical
chemical rate equations to \texttt{C} (or \texttt{CUDA}) code that,
among other things, implements these routines and generates the
corresponding reaction rate lookup tables in HDF5 format
\cite{hdf52018}.

We highlight the fact that the internal gas energy, $e_g$, may be
uniquely defined by the fluid fields, since
\begin{align*}
  e_t &= e_g + \frac{1}{2\rho}(m_x^2 + m_y^2 + m_z^2)\\
  \Leftrightarrow \quad &\\
  e_g &= e_t - \frac{1}{2\rho}(m_x^2 + m_y^2 + m_z^2),
\end{align*}
and thus this reaction network only adds 9 new independent fields to
the simulation.  However, due to the multirate solver structure
described in Section \ref{sec:implementation:time}, wherein
control over the simulation cleanly shifts between ``slow'' and
``fast'' phases, we use both $e_g$ and $e_t$ to store the
``current'' gas/total energy value.  Furthermore, by storing two
versions of the energy we may use our preferred \texttt{N\_Vector}
data structure layout -- five \texttt{N\_Vector\_Parallel} objects for
the fluid fields, plus one MPI task-local \texttt{N\_Vector} for the
chemistry fields.  We further note that this structure will enable
follow-on efforts in which the entire chemical network (data and
computation) are moved to GPU accelerators, through merely swapping
out our \texttt{N\_Vector\_Serial} object and enabling Dengo's
generation of GPU-enabled source code.

We initialize these simulations using a ``clumpy'' density field.
Here, the overall density field at any point $\bx\in\Omega$ is given
by the formula
\begin{equation}
  \label{eq:rho0}
  \rho(\bx,t_0) = \rho_0\left(1 + 5 e^{-20\|\bx-\bx_c\|^2} +
    \sum_{i=1}^{10\, n_p} s_i e^{-2(\|\bx-\bx_i\|/r_i)^2}\right),
\end{equation}
where $\rho_0=1.67\times 10^{-22}$ g / cm$^{3}$ 
is the background density, 
$n_p$ is the total number of MPI tasks in the simulation (i.e., the number of ``clumps'' is proportional to the number of MPI tasks),
and
$\bx_c = \left(\frac{x_l+x_r}{2}, \frac{y_l+y_r}{2}, \frac{z_l+z_r}{2}\right)$ 
is the location of the center of the computational domain.  We choose
the remaining parameters from uniform random distributions in the
following intervals: 
\begin{itemize}
\item $\bx_i \in \Omega$, i.e, each clump is centered randomly within
  the domain,
\item $r_i \in [3\Delta x, 6\Delta x]$ is the clump ``radius,''
  i.e., each clump extends anywhere from 3 to 6 grid cells away from 
  its center, and
\item $s_i \in [0, 5]$ is the clump ``size,'' i.e., each clump
  has density up to 5 times larger than the background density.
\end{itemize}

The simulation begins with a near-uniform temperature field, with only
a single higher-temperature region located in the clump at the domain
center:
\begin{equation}
  \label{eq:T0}
  T(\bx,t_0) = T_0\left(1 + 5 e^{-20\|\bx-\bx_c\|^2}\right),
\end{equation}
where the background temperature is chosen to be $T_0=10$ K.

The chemical fields are initialized to values proportional to the
overall density, with:
\begin{align}
  \label{eq:H2I0}
  \Hy_2(\bx,t_0) &= 10^{-12}\rho_0(\bx,t_0)\\
  \label{eq:H2II0}
  \Hy_2^+(\bx,t_0) &= 10^{-40}\rho_0(\bx,t_0)\\
  \label{eq:HII0}
  \Hy^+(\bx,t_0) &= 10^{-40}\rho_0(\bx,t_0)\\
  \label{eq:HM0}
  \Hy^-(\bx,t_0) &= 10^{-40}\rho_0(\bx,t_0)\\
  \label{eq:HeII0}
  \He^+(\bx,t_0) &= 10^{-40}\rho_0(\bx,t_0)\\
  \label{eq:HeIII0}
  \He^{++}(\bx,t_0) &= 10^{-40}\rho_0(\bx,t_0)\\
  \label{eq:HeI0}
  \He(\bx,t_0) &= 0.24\rho_0(\bx,t_0) - \He^{+}(\bx,t_0) - \He^{++}(\bx,t_0)\\
  \label{eq:HI0}
  \Hy(\bx,t_0) &= \rho_0 - \Hy_2(\bx,t_0) - \Hy_2^+(\bx,t_0) - \Hy^+(\bx,t_0)\\
  \notag
               &\quad- \Hy^-(\bx,t_0) - \He^+(\bx,t_0) - \He^{++}(\bx,t_0) - \He(\bx,t_0)\\
  \label{eq:de0}
  \el^-(\bx,t_0) &= \frac{\Hy^+(\bx,t_0)}{1.00794} + \frac{\He^+(\bx,t_0)}{4.002602} + 2\frac{\He^{++}(\bx,t_0)}{4.002602} - \frac{\Hy^-(\bx,t_0)}{1.00794} + \frac{\Hy_2^+(\bx,t_0)}{2.01588}\\
  \label{eq:ge0}
  e_g(\bx,t_0) &= \frac{k_b T(\bx,t_0) N(\bx,t_0)}{\rho(\bx,t_0) (\gamma-1)},
\end{align}
where $k_b=1.3806488\times 10^{-16}$ $\text{g}\cdot\text{cm}^2/\text{s}^2/K$ is Boltzmann's constant, 
$\gamma=\frac53$
is the ratio of specific heats, and the gas number density $N(\bx,t_0)$ is given by
\begin{align}
  \label{eq:Ndens}
  N(\bx,t_0) = 5.988\times 10^{23} \Bigg(
  & \frac{\Hy_2(\bx,t_0)}{2.01588} + \frac{\Hy_2^+(\bx,t_0)}{2.01588}
    + \frac{\Hy^+(\bx,t_0)}{1.00794} + \frac{\Hy^-(\bx,t_0)}{1.00794} \\
  & + \frac{\He^+(\bx,t_0)}{4.002602} + \frac{\He^{++}(\bx,t_0)}{4.002602}
    + \frac{\He(\bx,t_0)}{4.002602} + \frac{\Hy(\bx,t_0)}{1.00794} \Bigg).
\end{align}
Finally, we assume that the gas is initially static (i.e.,
$\mathbf{m}(\bx,t_0) = \mathbf{0}$), and that 
no external forces are applied (i.e., $\bG(\bx,t) = \mathbf{0}$).

We perform the simulations on the spatial domain $\Omega =
[0,3.0857\times10^{30}\: \text{cm}]^3$, and enforce reflecting boundary conditions
over $\partial\Omega$.  Our base grid simulations (discussed below)
are computed over the temporal domain $[0,10^{11}\: \text{s}]$.   

Since the Dengo-supplied chemical network expects values in CGS units,
but our FD-WENO solver prefers dimensionless quantities, we
non-dimensionalize by using the scaling factors: 
\begin{itemize}
\item \texttt{MassUnits} = $3\times 10^{70}\:\text{g}$,
\item \texttt{LengthUnits} = $3.0857\times 10^{30}\:\text{cm}$, and
\item \texttt{TimeUnits} = $10^{11}\:\text{s}$,
\end{itemize}
that correspond to the variable scaling factors
\begin{itemize}
\item \texttt{DensityUnits} = $1.0211\times 10^{-21}\:\text{g}/\text{cm}^3$,
\item \texttt{MomentumUnits} = $3.1507\times 10^{-2}\:\text{g}/\text{cm}^2/\text{s}$, and
\item \texttt{EnergyUnits} = $9.7223\times 10^{17}\:\text{g}/\text{cm}/\text{s}^2$,
\end{itemize}
which we use to convert between ``dimensional'' and ``dimensionless''
values as control is passed back-and-forth between fluid and chemistry
solvers.  Furthermore, we note that these choices of \texttt{LengthUnits}
and \texttt{TimeUnits} result in the normalized space-time domain
$[0,1]^3\times[0,1]$ -- all discussion of time step sizes or the temporal 
domain in the remainder of this report refer to these quantities in 
dimensionless units.

As is typical for large-scale explicit simulations, we perform weak
scaling studies by examining the time required \emph{per slow step}.
Thus as we refine the spatial discretization (e.g., $\Delta x \to
\frac{\Delta x}{2}$) and thus reduce the slow step size (e.g., $h^S
\to \frac{h^S}{2}$), we shorten the overall time interval for each
simulation (e.g., $[0,1]\to\left[0,\frac12\right]$), thereby
guaranteeing a steady number of ``slow'' time steps.  However, since
the fast time scale calculations are performed implicitly and thus
have no resolution-based CFL stability restriction, we may choose
between two alternate options: 
\begin{itemize}
\item[(a)] maintain an essentially-constant fast step size (i.e., $h^F
  \to h^F$), as would likely occur for chemical accuracy
  considerations alone, 
\item[(b)] refine the fast step size in proportion to the slow step
  size (i.e., $h^F \to \frac{h^F}{2}$), resulting in an essentially
  constant amount of work per MPI task per slow time step as the mesh
  is refined. 
\end{itemize}
Since the purpose of this report is to examine the parallel
scalability of the \texttt{MPIManyVector} and other general SUNDIALS
enhancements, we chose option (b) above, since option (a) would result 
in a decreasing number of fast time steps per simulation.  Thus the 
ensuing scalability results may be seen as a ``worst-case'' 
performance metric for scalability of multirate methods applied to 
physical problems of similar structure.   

We perform a standard explicit method weak scaling study for this
problem, wherein we increase the total number of nodes and total
problem size proportionately.  For each spatial mesh, we decrease the
``slow'' step size $h^S$ to maintain a constant CFL stability
factor (i.e., $h^S \propto \min(\Delta x, \Delta y, \Delta z)$).  In
each simulation, the initial ``transient'' phase runs over the time
interval $(0,0.1]$, and the ``fixed-step'' phase runs for the remainder 
of the time interval.  We set a value $h^F$ to
maintain a constant timescale separation ratio $h^S/h^F = 1000$; for the
transient phase $h^F$ this is the maximum step size for the temporal
adaptivity approach; $h^F$ then becomes the fixed step size for the 
latter portion of the calculation.  As stated above, we maintain an
essentially-constant computational effort per MPI task by shortening
the overall simulation time proportionately with $h^S$.  We note that
this causes the relative fraction of ``fixed-step'' versus
``transient'' portions of the simulation to decrease at larger scales.
However, since we use the same $h^F$ value for the fixed-step phase 
of the calculation as we use for the \emph{maximum} allowed transient 
time step size, then as the spatial mesh is refined we find  
that an increasing fraction of transient chemical time steps use this 
maximum value, thereby providing a robust solution approach that in the 
limit achieves a fixed amount of work per MPI task as the mesh is refined.  
In Table \ref{tab:scaling_parameters} we provide a
summary of these parameters for each problem size tested.
\begin{table}[!htp]
  \begin{center}
    \begin{tabular}{|r|c|c|c|c|l|}
      \hline
      Nodes & Spatial Mesh & Total Unknowns & Slow step & Fast step & Final time \\
      \hline
         2 &    $125 \times 100 \times 100$ & $1.875\times 10^{7}$ & $1.00\times 10^{-1}$            & $1.00\times 10^{-4}$            & $1$\\
        16 &    $250 \times 200 \times 200$ & $1.5\times 10^{8}$ & $5.00\times 10^{-2}$            & $5.00\times 10^{-5}$            & $0.5$\\
       128 &    $500 \times 400 \times 400$ & $1.2\times 10^{9}$ & $2.50\times 10^{-2}$            & $2.50\times10^{-5}$             & $0.25$\\
       432 &    $750 \times 600 \times 600$ & $4.05\times 10^{9}$& $1.6\overline{6}\times 10^{-2}$ & $1.6\overline{6}\times 10^{-5}$ & $0.16\overline{6}$\\
      1024 &   $1000 \times 800 \times 800$ & $9.6\times 10^9$ & $1.25\times 10^{-2}$            & $1.25\times 10^{-5}$            & $0.125$\\
      2000 & $1250 \times 1000 \times 1000$ & $1.875\times 10^{10}$ & $1.00\times 10^{-2}$            & $1.00\times 10^{-5}$            & $0.1$\\
      3456 & $1500 \times 1200 \times 1200$ & $3.24\times 10^{10}$ & $8.3\overline{3}\times 10^{-3}$ & $8.3\overline{3}\times 10^{-6}$ & $0.083\overline{3}$\\
      \hline
    \end{tabular}
    \caption{Weak-scaling simulation parameters.  Each simulation uses
      40 CPU cores per node on Summit \cite{Summit_site}, i.e., these
      simulations ranged from 80 to 138,240 CPU cores.  All simulations
      begin with an initial time of $t_0=0$, run the ``transient''
      phase over the time interval $(0,\min(0.1,t_f)]$, and run the
      ``fixed-step'' phase over any remaining interval, $(\min(0.1,t_f),t_f]$. 
      \label{tab:scaling_parameters}
    }
  \end{center}
\end{table}

For each spatial mesh size, we performed two simulations.  The first
used two ``newer'' features added to the SUNDIALS \texttt{N\_Vector}
API: 
\begin{itemize}
\item \emph{Fused vector operations} -- for back-to-back
  \texttt{N\_Vector} operations that are performed in SUNDIALS'
  various solvers, we created a set of new \texttt{N\_Vector} kernels
  that perform this multiple combination of operations per memory
  access, thereby increasing arithmetic intensity and reducing the
  number of function calls (or GPU kernel launches). 
\item \emph{Local reduction operations} -- since an
  \texttt{MPIManyVector} is merely a vector that is comprised of a
  subset of other \texttt{N\_Vector} objects, any operation requiring
  an MPI reduction (e.g., dot-product or norm) would naively
  result in multiple separate \texttt{MPI\_Allreduce} calls.  
  We therefore created a set of new \texttt{N\_Vector}
  routines that only perform the MPI task-local portion of these
  operations, waiting to call \texttt{MPI\_Allreduce} until the
  overall accumulated quantity is available, and thereby reducing MPI
  latency effects. 
\end{itemize}
The second simulation performed at each spatial mesh size was run with
these newer features disabled, so that we could assess the benefits of
these newer features ``at scale.''

\section{Performance Results}
\label{sec:performance}

For each simulation we manually instrumented the code with 
profiling timers based on \texttt{MPI\_Wtime} for a variety of 
logical subsets of the code: 
\begin{itemize}
\item[(a)] Setup -- this includes construction of \texttt{MRIStep},
  \texttt{ARKStep}, and \texttt{Dengo} data structures, as well as
  construction of initial condition values.  We note that due to the
  sum over $10 n_p$ clumps in creation of the initial density field
  \eqref{eq:rho0}, this component should exhibit slowdown as the
  problem size is increased.  We therefore do not include this 
  contribution when assessing the overall parallel efficiency. 
\item[(b)] I/O -- this includes all time spent in output of diagnostic
  information to \texttt{stdout} and \texttt{stderr} as the simulation
  proceeds, as well as all time spent in reading and writing HDF5
  files.  For the results that follow, most of these capabilities were
  disabled so that disk I/O would not affect our performance measurements. 
\item[(c)] MPI -- this includes all time spent in parallel
  communication \emph{by this demonstration application code}.  This 
  \emph{does not} include the MPI time spent in SUNDIALS'
  \texttt{N\_Vector} operations (e.g., dot-products and norms). 
\item[(d)] Packing -- this includes all time spent in packing the
  contiguous memory buffers (see steps 2 and 4 from Section
  \ref{sec:implementation:space}) that are used in the FD-WENO flux
  reconstruction. 
\item[(e)] FD-WENO -- this includes all time spent in performing the
  FD-WENO flux reconstruction. 
\item[(f)] Euler -- this includes the time spent in all steps from
  Section \ref{sec:implementation:space}, i.e., it should include the
  sum of (c)-(e) above, as well as the time required to evaluate the
  flux divergence \eqref{eq:div_flux}. 
\item[(g)] \texttt{fslow} -- this includes the time spent in (f)
  above, as well as all translation when converting between
  dimensional and dimensionless units for the fluid and chemical
  fields. 
\item[(h)] \texttt{ffast} -- this includes only the time spent in
  evaluating the ODE right-hand side for the Dengo-supplied chemical
  network. 
\item[(i)] \texttt{JFast} -- this includes only the time spent in
  evaulating the Jacobian of \texttt{ffast}, and storing those values
  in CSR matrix format. 
\item[(j)] Linear solver setup -- this includes all time spent by our
  custom \texttt{SUNLinearSolver} implementation to factor its 
  block-diagonal linear system matrices; this is essentially just the 
  time spent in MPI task-local calls to the \texttt{SUNLinSol\_KLU} 
  ``Setup'' routine. 
\item[(k)] Linear solver solve -- this includes all time spent by our 
  custom \texttt{SUNLinearSolver} implementation to solve its block-diagonal 
  linear systems; it is essentially just the time spent in MPI task-local 
  calls to the \texttt{SUNLinSol\_KLU} ``Solve'' routine. 
\item[(l)] Overall ``transient'' simulation time -- this includes all
  time spent in evolving the solution over the initial transient time
  interval, $(t_0,t_0+h_t]$.  We note that this \emph{does not} include the
  ``Setup'' time (a) above. 
\item[(m)] Overall ``fixed-step'' simulation time -- this includes all
  time spent in evolving the solution over the second fixed-fast-step
  time interval, $(t_0+h_t,t_f]$. 
\item[(n)] Total simulation time -- this includes \emph{both} the
  ``transient'' and ``fixed-step'' simulation times, as well as the
  ``Setup'' time for the simulation, i.e., $(a+l+m)$. 
\end{itemize}
We 
note that although these profilers only directly measure 
the time spent in the ``multiphysics'' portions of this simulation, we
may indirectly measure the overall amount of time spent in SUNDIALS
modules (\texttt{MRIStep}, \texttt{ARKStep}, \texttt{N\_Vector}, etc.)
by subtracting the amount of time spent in right-hand side, Jacobian,
and linear solver operations from the overall amount of time spent in
evolving the system, i.e.,
\[
  \text{SUNDIALS time} = (l + m) - (g + h + i + j + k)
\]
We further note that the vast majority of synchronization points
in this code occur in \texttt{N\_Vector} reduction operations (e.g., 
dot-products and norms), and thus the ``SUNDIALS time'' includes all 
such synchronization points.  While some local synchronization occurs from 
nearest-neighbor communication for flux calculations, this is overlaid 
by flux computations over subdomain interiors, and is directly measured 
by the timers (c), (f), and (g) above. 

\begin{figure}[htb!]
    \centering
    \includegraphics[width=0.85\textwidth]{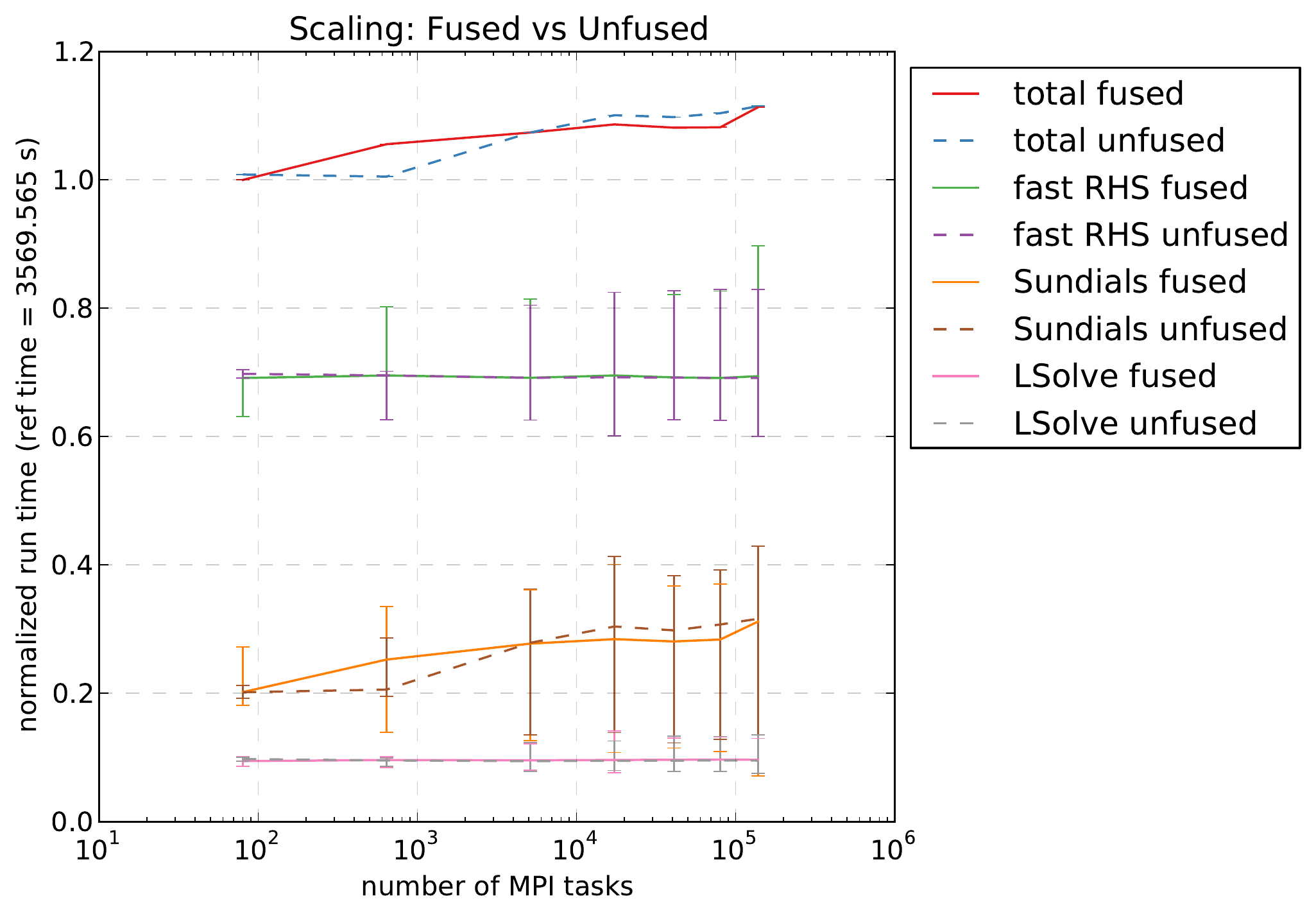}
    \caption{Weak scaling results for both fused and unfused \texttt{N\_Vector}
    versions.  Only the code components requiring 
    over 0.5\% of the total runtime are shown.  The parameters used for these tests 
    (grid sizes, nodes, etc.) 
    are provided in Table \ref{tab:scaling_parameters}.}
    \label{fig:scaling_compare}
\end{figure}

We provide weak-scaling results with these timers in Figure
\ref{fig:scaling_compare}.  While we collected 
data on a large number of code components, most of those timers required only 
trace amounts (less than 0.5\%) of the total runtime: Setup, I/O, MPI, Packing, 
FD-WENO, Euler, \texttt{fslow}, \texttt{Jfast}, and linear solver setup.  We 
have thus removed these measurements from this figure to 
focus on the time-intensive aspects of the 
code, as well as their parallel scalability.  Since each profiler resulted in 
slightly different times on each MPI task, for each curve we plot both the mean 
task time, as well as error-bars showing the minimum and maximum reported 
values.  However, since the `Sundials' times are only measured indirectly, 
these minimum and maximum values may be exaggerated since they incorporate the 
variability from other portions of the code (in addition to variability 
resulting from within SUNDIALS itself).  Thus the error bars for the `Sundials' curve in this Figure likely over-estimate the variability encountered by MPI tasks in the SUNDIALS infrastructure.
Additionally, in Table \ref{tab:efficiency} we provide the parallel efficiency of each simulation, in comparison with the 80-task ``fused'' simulation.
\begin{table}[!htp]
  \begin{center}
    \begin{tabular}{r|ccccccc}
         & \multicolumn{7}{c}{MPI tasks}\\
         & 80 & 640 & 5,120 & 17,280 & 40,960 & 80,000 & 138,240 \\
      \hline
      Fused & 1.00 & 0.95 & 0.93 & 0.92 & 0.92 & 0.92 & 0.90 \\
      Unfused & 0.99 & 0.99 & 0.93 & 0.91 & 0.91 & 0.90 & 0.90 \\
    \end{tabular}
    \caption{Parallel efficiency of weak-scaling simulations, for both ``fused'' and ``unfused'' \texttt{N\_Vector} versions.
      \label{tab:efficiency}
    }
  \end{center}
\end{table}

We first note that at the smallest scale tested ($125\times 100\times 100$ grid
with 80 MPI tasks), the fast chemical right-hand side routine accounted 
for almost 70\% of the total runtime, followed by general SUNDIALS infrastructure 
(slightly over 20\%), and linear system solves (slightly under 10\%).  
Furthermore, of these the fast 
chemical right-hand side routine and linear solver times showed perfect 
weak scalability, while the time spent in SUNDIALS infrastructure increased 
by almost 42\%, leading to run-time increase of about 11\% for the overall simulation 
code (i.e., approximately 90\% parallel efficiency compared to the smallest test).  
We additionally note that the variance 
in reported times increased at larger scales, particularly for the SUNDIALS
infrastructure, that showed up to 77\% variance at 138,240 MPI tasks.  Interestingly,
the various MPI tasks showed essentially zero variability in the total simulation times 
reported.  This is likely due to the fact that these ``total simulation'' timers surrounded calls to 
\texttt{MRIStepEvolve}; since the last thing this does before returning is compute an 
error norm (via \texttt{MPI\_Allreduce}), it effectively synchronizes all MPI tasks just 
prior to stopping the timer.
Finally, we note that the ``fused'' and ``unfused'' versions showed no
statistically-significant differences.

\section{Conclusions}
\label{sec:conclusions}

While we expected the general strong performance results shown in Section
\ref{sec:performance}, some of these came as more of a surprise than others.

From an application viewpoint, we note that our approach for interleaving 
communication and computation for the advective fluxes proved sufficient, with 
all  directly-measured ``MPI'' timings remaining at essentially zero for all 
scales tested.  Moreover, due to the multirate structure of this application
problem and testing setup, the advective portion of the right-hand side was 
essentially ``free'' in comparison to the chemical kinetics that dominated 
the simulation time.

From a SUNDIALS-infrastructure viewpoint, we posit that the slowdown shown in
the SUNDIALS infrastructure was entirely due to synchronization-induced 
latency in \texttt{MPI\_Allreduce} calls.  This conclusion is based on two 
observations.  First, while the ``fused'' version reduced the \emph{number} 
of these calls by 80\%, it left the \emph{frequency} of these synchronization
points essentially unchanged, and thus the near-identical runtimes for both 
indicate that the additional \texttt{MPI\_Allreduce} calls in the ``unfused'' 
version do not increase the overall global synchronization of the code.  
Moreover, the large variance reported 
for SUNDIALS infrastructure among MPI tasks indicates that some tasks spent 
considerably more time waiting at these synchronization points than others.
However, since we could only indirectly measure the time spent in SUNDIALS 
routines, this variance may be exaggerated.

Based on the above, we anticipate that this code will tremendously benefit 
from three planned investigations.  First, we plan to extend this implementation to
construct a custom \texttt{SUNNonlinearSolver} for the fast time scale.  As 
discussed in Section \ref{sec:implementation:algebraic}, this can be constructed 
to remove all \texttt{MPI\_Allreduce} calls at the algebraic solver level, in 
turn reducing over 98\% of the overall \texttt{MPI\_Allreduce} calls from the 
``fast'' time scale (for the 138,240 MPI task run, there were 61,195 Newton 
iterations and 1,007 fast time steps; $61195/(61195+1007) \approx 0.984$), thereby 
effectively removing nearly all global synchronization from these simulations.

Second, we plan to port the chemical network data, the fast right-hand side 
computations, the fast Jacobian construction routine, and the fast-time scale 
algebraic solvers to GPU accelerators.  
We note that even at the largest scales tested here
($1500 \times 1200 \times 1200$ grid on 138,240 MPI tasks), evolution 
of these arithmetically-intense ``fast'' chemical processes required over 
70\% of the total runtime (fast RHS + lsolve).  Moreover, 
the types of calculations performed in this module should be amenable to GPU
architectures, and thus their conversion could likely result in 
\emph{significant} performance improvements over the results shown here.

Third, instead of requiring application codes to only indirectly measure
the time spent in SUNDIALS, and thus lumping all of SUNDIALS' various
actions (\emph{plus} any unmeasured work) into a single performance metric, 
future performance studies with SUNDIALS would benefit tremendously from 
\emph{direct} measurements of SUNDIALS performance, notably if these 
measurements were broken apart into logical units (e.g., vector reductions, 
vector arithmetic, integrator infrastructure, nonlinear solvers, linear 
solvers, preconditioners, etc.).  We are thus investigating inclusion of 
direct measurements of SUNDIALS performance through a tool such as Caliper
\cite{caliper_site}.  Additionally, such tools could track more than just 
runtime, enabling measurement of system-level counters.

\section*{Acknowledgements}
This research was supported by the Exascale Computing Project (ECP), Project Number: 17-SC-20-SC, a collaborative effort of two DOE organizations – the Office of Science and the National Nuclear Security Administration, responsible for the planning and preparation of a capable exascale ecosystem, including software, applications, hardware, advanced system engineering and early testbed platforms, to support the nation’s exascale computing imperative.

Support for this work was also provided through the Scientific Discovery through Advanced Computing (SciDAC) project ``Frameworks, Algorithms and Scalable Technologies for Mathematics (FASTMath),'' funded by the U.S. Department of Energy Office of Advanced Scientific Computing Research and National Nuclear Security Administration. 

This research used resources of the Oak Ridge Leadership Computing Facility, which
is a DOE Office of Science User Facility supported under Contract DE-AC05-
00OR22725.

This work work was performed under the auspices of the U.S. Department of Energy by Lawrence Livermore National Laboratory under contract DE-AC52-07NA27344. Lawrence Livermore National Security, LLC.  LLNL-TR-791538.

This document was prepared as an account of work sponsored by an agency of the United States government. Neither the United States government nor Lawrence Livermore National Security, LLC, nor any of their employees makes any warranty, expressed or implied, or assumes any legal liability or responsibility for the accuracy, completeness, or usefulness of any information, apparatus, product, or process disclosed, or represents that its use would not infringe privately owned rights. Reference herein to any specific commercial product, process, or service by trade name, trademark, manufacturer, or otherwise does not necessarily constitute or imply its endorsement, recommendation, or favoring by the United States government or Lawrence Livermore National Security, LLC. The views and opinions of authors expressed herein do not necessarily state or reflect those of the United States government or Lawrence Livermore National Security, LLC, and shall not be used for advertising or product endorsement purposes.

\bibliographystyle{siam}
\bibliography{refs}

\end{document}